\documentclass[3p,times,twocolumn]{elsarticle}
\usepackage{graphicx}
\usepackage{ecrc}

\volume{00}

\firstpage{1}

\journalname{Nuclear Physics B Proceedings Supplement}

\runauth{D. Fargion, P. Oliva}

\jid{nuphbp}

\jnltitlelogo{Nuclear Physics B Proceedings Supplement}

\usepackage{amssymb}

\usepackage[figuresright]{rotating}
\usepackage{amsmath}
\usepackage{xspace}

\begin{document}

\begin{frontmatter}



\dochead{}

\title{Crossing muons in Icecube at highest energy: \\
a cornerstone to {\boldmath $\nu$} Astronomy}


\author{D. Fargion$^a$ \& P. Oliva$^b$}

\address{$^a$Physics Department and INFN, Rome University 1, Sapienza, Ple. A. Moro 2, Rome Italy}
\address{$^b$Niccol\`o Cusano University, Via Don Carlo Gnocchi, 3, 00166 , Rome, Italy}

\begin{abstract}\noindent
Highest energy  neutrino events (contained) in cubic km ICECUBE detector
resulted
in last three years
to be
as many as $37-2=35$  signals (two of those
having
been recently discharged); these tens-hundred TeV (32 energetic events) up to
rarest
(only
3) PeV cascade showers,
proved to have
an extraterrestrial
origin.
Their flux exceeded, indeed, the expected atmospheric noise
and clearly favored and tested the birth of a long waited $\nu$ astronomy.
The UHE neutrino flavor transition from a $\nu_{\mu}$ atmospheric dominance (over $\nu_{e}$ showers at TeV energy), toward a higher energy shower cascade ($\nu_{e}$, $\nu_{\tau}$) events at few tens TeV up to PeV energy is a hint of such a fast extraterrestrial injection.
The majority (28
out of 35)
of all these events are spherical cascade showers and their exact timing in shower shining provided an approximate $\nu$
arrival direction, within
about $\pm10^{\circ}$.
However, their consequent smeared map
is
inconclusive: both because of such a wide angle spread signal of $\pm10^{\circ}$ and
because of their paucity, is not yet allowable to define any meaningful source correlation or anisotropy.
The additional rarest $9-2=7$ muon tracks, while being sharp in arrival directions, did not offer any correlated clustering nor any
overlapping within known sources.
Larger sample of UHE $\nu$ signals and their most accurate directionality is needed. We recently suggested that the highest energy
(tens-TeV) crossing muon along the ICECUBE, mostly at horizons or upcoming, are the ideal tool able to reveal soon such clustering or even any narrow angle pointing to known (IR, X, Radio or $\gamma$) sources or self-correlation in rare doublet or triplet: a last hope for a  meaningful and short-time $\nu$ Astronomy. Any crossing  muons clustering along galactic sources or within UHECR arrivals might  also probe rarest (possibly galactic, radioactive and in decay in flight) UHECR  event made by nuclei or neutrons. Within three years of ICECUBE data all the non-contained crossing highest energy muons above few tens TeV may be several dozens, possibly around $54$, mostly enhanced along horizontal edges, painting known sources and/or self-correlating in doublets or rarest triplet, offering a first solution of the UHE neutrino source puzzle (if steady or transient nearby source are at sight). Recent preliminary ICECUBE presentation on crossing muons are consistent with our preliminary muon rate estimate.
\end{abstract}


\end{frontmatter}
\section{ Neutrino Astronomy in ICECUBE}
The presence of Cosmic Rays, CR, their sources and their acceleration is presently an
open problem in high-energy astrophysics. Cosmic rays are able to be accelerated because CR are charged particles.
Unfortunately for the same reason CR charges suffer of relic (large scale) galactic and extragalactic magnetic field bending.
Such a random walk in the magnetic field forest make smooth and homogeneous their arrival direction.
Incidentally the same presence of such large scale magnetic fields  test the absence
in CR of any detectable magnetic monopole particle trace, the so called Parker bound;
we know of such far magnetic fields presence by the consequent Faraday rotation of far polarized  radio-sources.
Therefore CR are smeared and do not offer any CR astronomy.
The rarest Ultra High Energy Cosmic Rays, UHECR, above tens EeV,
 were expected to be less bent and to correlate with their
 nearby (Super-Galactic) sources because of their rigidity and straight directionality. This hope rose few years ago and it faded quite soon.
 No super-galactic imprint in UHECR maps has been found yet.
 Indeed, an additional variable generate confusion and smearing: the UHECR composition
 has been observed as heavy nuclei (AUGER) or light nuclei or protons (Hires-TA). Therefore CR are smeared.
For a comparable reason the smeared CR while hitting the Earth atmosphere mimic the mess
by producing a diffused rain of secondaries pions $\pi^{\pm}$, kaons $K^{\pm}$ and muons $\mu^{\pm}$, whose final traces
in underground detectors are also smeared neutrinos: the so called atmospheric neutrinos.
Therefore any eventual neutrino astronomy is drowned in such a smooth sea of
atmospheric neutrino noise. Neutrinos have their own identities, or flavors: they do not behave at same way.
In effect, the slow decay of muons respect to the $\pi^{\pm}$ one or the $K^{\pm}$ one, makes above few tens-hundred GeV the
atmospheric $\nu_{e}$ flux more rare respect to the $\nu_{\mu}$ flux nearly by an order of magnitude.
This implies a muon-rich signal at TeVs (long track traces) respect to rarer $\nu_{e}$ showers observed in Deep Core
inside ICECUBE as small cascade showers.
Therefore as soon as ICECUBE highest energy events have shown
ruling cascades (mostly originated by  $\nu_{e}$ or $\nu_{\tau}$ charged current, CC, interaction), then
the atmospheric neutrino flux \cite{Gaisser02} \cite{Enberg08}, has been overcome by a new neutrino sky, mostly of extraterrestrial and astrophysical nature.

Originally the ICECUBE attention was for the search of UHE neutrinos at EeV GZK cosmological edges \cite{za66}, but recent results are at lower PeV energy windows. Tau EeV neutrinos  $\nu_{\tau}$, $\bar{\nu_{\tau}}$ might hit the Earth, produce and EeV $\tau$ lepton whose escape and decay in flight becoming observable (and searchable) as an horizontal fluorescence $\tau$ airshower \cite{Fargion02}; this probable event has not been observed yet \cite{AUGER13}.  Neutrino oscillation \cite{Fargion-2012} and mixing \cite{Fargion-2011} from far galactic or extragalactic distance may overshadow most atmospheric neutrino flavor composition ruled by a final flavor ratio at TeV: $(\nu_{e}, \nu_{\mu}, \nu_{\tau})\div\left(\frac{1}{10}, 1, 0\right)$,
into a more ``democratic" flavor composition above 30 TeV, as the observed one, approximately of $(\nu_{e}, \nu_{\mu}, \nu_{\tau})\div(1, 1, 1)$,  also assuming the mild additional presence of neutral current cascades \cite{Vissani-2013}, \cite{DFPP14}.
These signals might be born by a huge AGN flaring jets or by more abundant GRB precessing jets in competition with
  more conventional SNRs-microjets  sources possibly origin of CR at lower (PeVs) energy edge. These extra-terrestrial events may be both of galactic and extragalactic nature.

The CR and UHECR neutral parasite secondaries, $\gamma$, X, radio synchrotron signals suggest the
AGN, BL Lac hypothesis for UHE $\nu$. The one-shoot GRB model is not well correlated up to our days with
 any observed UHE neutrino in ICECUBE. Some rare precursor event a few hour before the $\gamma$ burst might be correlated,
 but they call for a long life precessing gamma jet
model \cite{Fargion99} often ignored respect to the (still) popular one shoot fireball model.
 More common AGN, Galactic Cluster, star forming clusters  or extragalactic IR sources are the possible birth place of UHE neutrinos.
Therefore we need a better view of the CR and possibly their related inner probe made by UHE neutrinos.

 UHECR were expected to produce (by scattering on BBR photons) an observable rate of photo-pions and EeV neutrino
 (the cosmogenic neutrinos). This scattering on relic photons lead to an opacity, the so-called GZK cut off in UHECR spectra, that is still experimentally unsettled because it might be in debt also
 of an intrinsic acceleration limit and/or to a changing mass composition role.
 Therefore the PeV neutrinos are not clearly related to such GZK EeV UHECR. \cite{DFPP14}. These UHECR cut off in GZK opacity, \cite{za66}, the consequent cosmo-genic neutrinos are possibly better observable soon as Tau airshower \cite{Fargion02}
 at EeV (also so called  Earth-Skimming neutrinos \cite{Feng02}) in AUGER \cite{Fargion02}, \cite{Bertou2002}, HIRES, TA or ASHRA array telescopes; such a Tau airshower signal has not been yet revealed, although the time seem already mature, at least in ASHRA \cite{Aita2011} experiment tuned to PeV energies. The Tau airshower astronomy is a secondary tail of the \cite{doub_bang} Double Bang proposal, that might be observable also in ICECUBE by ellipsoidal or separate PeVs future events.

However, in conclusion, the  severe $\gamma$ BBR opacity to photons above galactic distances at PeV (10 kpc), suggest PeV neutrino of extragalactic origin. Let us remind that the recent ICECUBE spectrum near PeV is tailed and cut \cite{DFPP14} to avoid any (enhanced and also expected) resonant $\bar{\nu_{e}} + e\rightarrow W^{-}$ event at $6.3$ PeV \cite{glashow}.  Therefore the novel  extraterrestrial signal at PeV \cite{Science-2013} and below is fine-tuned to be suppressed at higher energies. As we mentioned in the introduction the contained events are mostly cascade showers whose angular resolution is poor: thus, any correlation with sources or other mass distribution become difficult.
 The absence of high angular resolution for cascades and the rareness (7) of contained $\nu_{\mu}$ events makes
 the ability to radically improve  such contained UHE neutrino astronomy critical.

\section{Contained versus crossing muons at tens TeVs-PeV}

The high energy muon $\mu$ , $\bar{\mu}$ are the most penetrating muons (up to EeV energy where
$\tau$ , $\bar{\tau}$ become the winning leptons) and they may be originated well outside the same
ICECUBE volume. Their larger size detection simply amplify the $\nu_{\mu}$ neutrino volume and their presence: the energy losses of the muons is (within TeVs-PeVs energy) reasonable foreseen, growing proportionally to the muon energy,
therefore linking their emission photon number to their energy and its $\mu$ length to the  logarithmic energy growth; these distances are leading to a larger volume and a wider rate of crossing muon neutrino events.
 The muon distance in the water or ice may be described \cite{Fargion02} within (TeV-PeV) by this simple
 phenomenological law:
 \begin{equation}
 \centering
L_{\mu} \simeq L_{0}\cdot \left[1 + \frac{3}{2} \log\left(\frac{E_{\mu}}{\text{TeV}}\right)\right]
\end{equation}
Where $ L_{0} = 2.6$ km.  These large distances above tens TeV or PeV energy makes the effective detector for UHE
$\nu_{\mu}$ an order of magnitude larger than the other contained flavors (cascade-shower, bounded spherical events).
The most recent declination distribution spectra shown in Moriond \cite{C.Kopper2014}  for the recent $37$ ICECUBE events is allowing
us to extrapolate the expected crossing TeV muon number at each declination by simple approximations
(based on geometry and muon lenght). The volume and the rate is proportional to the
\emph{allowed muon track distance outside} the ICECUBE detector. The Earth distribution around the detector, the Earth opacity, the allowed muon distance in above formula around the $50$ TeV energy will cross almost $9$ km: one might imagine to multiply simply this number for the observed $8$ event reaching $72$
expected signals. This first value is quite over-estimated because the geometry is not equally spread around ICECUBE:
the mass above the detector is only $2$ Km depth. Moreover the observed angular distribution event rate has to been more carefully considered leading to smaller estimate. Rounding (to unity)  each zenith width spectra, from downward to upward as in figure,  we found  as shown in figure below an amplified number of crossing $\mu$ events (respectively from downward to upward versus: $3$, $3$, $5$, $7$, $8$, $8$, $7$, $6$, $4$, $3$), for each zenith arrival, for a total
of nearly $54$ (but a wide error margin below or about $50\%$). Let us try with present numbers an an early statistical forecast.

\section{Probability to find pairs, triplets in the $\mu$ up-going sky}
Let us estimate for present (approximated) $54$ forecast crossing UHE muons events, each of them occupying a tiny solid angle of area
$a$ (for  a muon resolution angle $\theta \simeq 1^{\circ}$), $a \simeq \frac{\pi}{57^{2}}\simeq 9.7 \cdot 10^{-4}$ sr respect
the whole $4\cdot \pi$ of the sky, is  $\frac{a}{4\pi} =$$ \frac{\Delta \Omega}{\Omega}$. This  ratio, $\varepsilon = \frac{a}{4\pi}$ is very small indeed:
 $$ \varepsilon \simeq \frac{\pi \cdot \theta^{2}}{4 \pi} \simeq 10^{-4}$$
 The last enhanced approximation takes place because we considered (again as a zero order approximation)
 at $30-50$ TeV energy, a partial neutrino opacity  to the Earth
 reducing the whole observable sky from $4\pi$ to nearly $3\pi$.
Therefore the probability to overlap with any other different neutrino event may be estimated.
 It is more convenient to first estimate the probability $\neg P$ \emph{not to observe}
  any overlapping (doublet, triplet or poker) in the sky; this value being:
 \begin{equation}
\neg P= 1\cdot(1-\varepsilon)\cdot(1-2\varepsilon)\cdots[1-(n-1)\varepsilon];
\end{equation}
This  product  sequence might be estimated either imagining an average quadratic expression or, in equivalent way,
 considering first $P = e^{\ln P}$, secondly by making the exponent product sequence as a sum of each $\ln(1-j\varepsilon)$ term ($\forall j=1\cdots n-1$);  because of the tiny value of the upper bound  ($n \cdot \varepsilon \leq 5.4 \cdot 10^{-3}$) each  logarithmic element maybe approximated to
  $\ln(1-j\varepsilon)\simeq  j\varepsilon$. In this way the whole exponent sum becomes:
\begin{equation}
\neg P=  e^{\sum \ln(1-j\varepsilon)}\simeq  e^{-\varepsilon\sum j} = e^{-\varepsilon \frac{n(n-1)}{2}}
\end{equation}
  This value assuming our  (tentative) estimate of muon crossing in three years is:
  \begin{equation}
\neg P=  e^{-0.143} \simeq 0.867
\end{equation}
Therefore the probability to \emph{not} find pairs is large, and
the consequent probability to observe at least a doublet (or more) is $1-(\neg P)\simeq0.133$, being 13\% low but not extremely low.

The probability to discover at least two pairs could be smaller and
we estimated it, once again assuming
as a first approach   the product of the same
probability to observe at once twice of such an event:  ${P_{\geq 2} \simeq [1-(\neg P)]^{2}= 1.77\%}$.
 To be more correct one may estimate the same probability to observe at least two pairs as
 certainty (1) minus the probability of no pair ($P_0$) or just one pair ($P_1$).
 The result (comparable to the previous one) is:
 $$
 P_{\geq 2}= 1- P_{0}-P_{1}\simeq
 $$
    \begin{equation}
\simeq \left[1- e^{-\varepsilon \cdot \frac{n\cdot(n-1)}{2}}\left(1 +\varepsilon\cdot \frac{(n-1)(n-2)}{2}\right)\right]
\end{equation}
This result in such a first forecast may be also written as follows:
  \begin{equation}
P_{\geq 2}\simeq \left[\varepsilon\cdot\frac{(n-1)^{2}}{2}\right]^{2} = 1.97\%
\end{equation}

Therefore the finding of two or more pairs may reduce the probability to be a chance
(or viceversa it may confirm the $\nu$ self-correlation) at a small  percent. Three pairs  as
we may expect reduce this probability at nearly below $0.4\%$; the same presence of just one triplet may also
reduce the probability to be a chances as small as  $P_{3} \simeq 0.07\%$. A quadruplet or more multiplets
might drive the neutrino astronomy even to a potential test of flavor ratio estimates, a revolution now, beyond to our
most optimistic dreams.

\begin{figure}[h]
\includegraphics[scale=0.33]{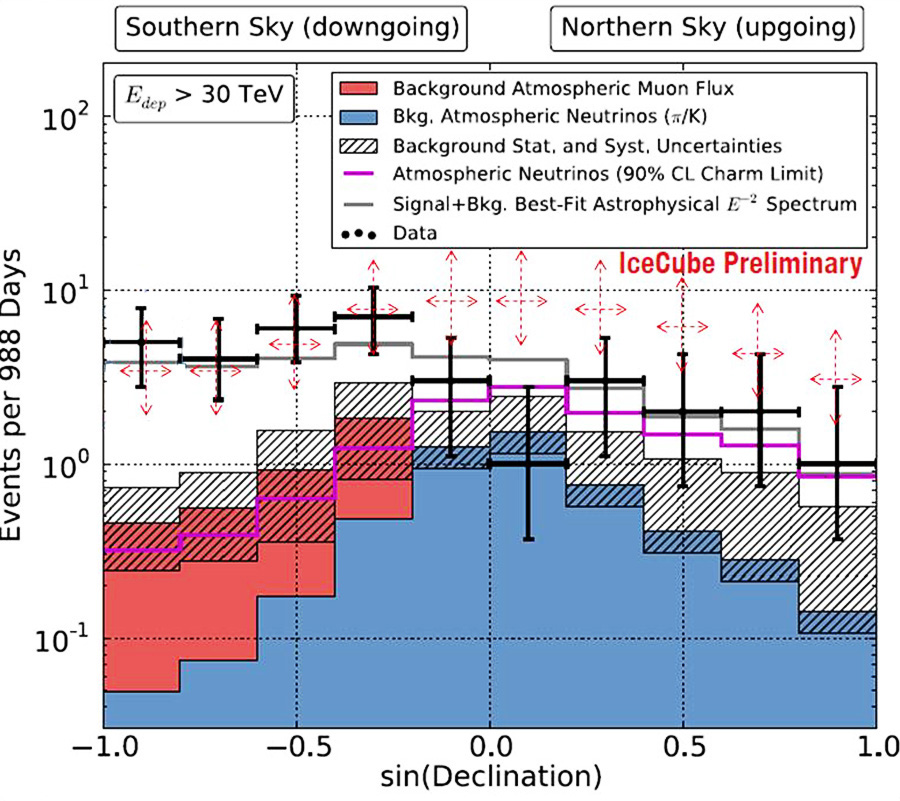}
\caption{The  dashed thin red crosses foresee the crossing muons numbers for three years in ICECUBE, assuming an energy threshold about $30-50$ TeV. Their total large (nearly $54$) number and they track narrow beam may dig in the sky map
correlations and multiplet clustering along known sources.} \label{Foreseeing}
\end{figure}

\section{Conclusions and updates}
\begin{figure}[h]
\includegraphics[scale=0.33]{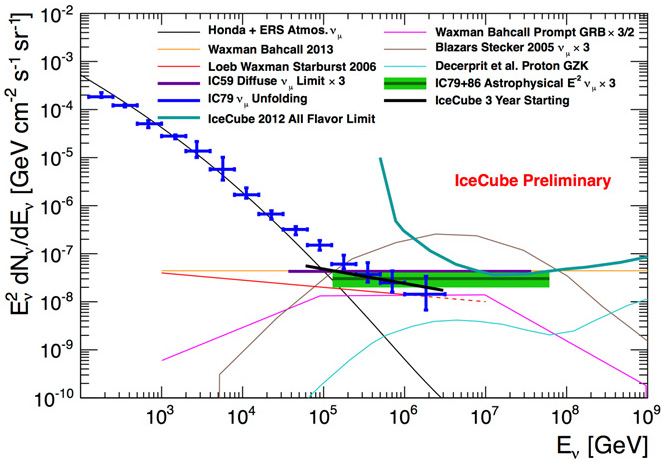}
\caption{The preliminary flux of crossing muons delivered only on second June 2014 at Boston-Neutrino14; their horizontal and upgoing rate in two years is consistent with the fraction of up-going foreseen in  \cite{Fargion-2014}, (nearly $30-40$). } \label{Crossing}
\end{figure}

The discover of the highest energy neutrino astronomy require a high resolution probe.
Recent $37-2=35$ ICECUBE highest energetic events are mostly $28$ cascade showers with poor arrival angle $\pm 10^{\circ}$, while
 $7$ muon tracks directions for contained events point to a much narrow angle as  $\pm 1^{\circ}$, but they are rare; therefore muon track solid angle is more than two order of magnitude smaller and sharper by solid angle than  cascade ones. The present spread shower signals in ICECUBE maps are not useful to address clearly to any smoking gun sources, nor to test large scale anisotropy.
 Clustering along a source,  possibly along galactic regions (in analogy to the  observed  Cen-A UHECR multiplet events  in  AUGER maps or to the ARGO-MILAGRO anisotropy sky at TeVs CR), might favor also the presence of UHECR radioactive decay in flight, bent by magnetic fields \cite{Fargion-2011b}, whose decay secondaries could be $\gamma$ and also TeVs-PeV neutrinos.
 However the need for sharp neutrino maps is compelling.
The abundant muon crossing at highest (tens TeV) tracks, tagged by their huge energy release, are self selected as extraterrestrial and
they are a very rich key to discover by a  sharp view the highest energy neutrino  sky. A rare doublet may be not yet convincing; but two or above doublets and/or  rare triplet within the expected $60$ events may make the steps into  neutrino astronomy.
Very recent presentation \cite{G.Hill14} , see Fig. \ref{Crossing} did show the preliminary crossing muons spectra (but not their map yet)  corresponding to upgoing and horizontal crossing muons. Their rate for two years is nearly $40$ events, consistent with our recent \cite{Fargion-2014} and here reconfirmed  estimate.

\section{Acknowledgment}
The author wish to thank Prof. Filippo Cesi for his useful comments and help.
\section{Dedication}
This article is devoted to the memory of Dr. Ameglio Moshe Fargion,
(2-06-1913 -- 22-04-2011), Assistent Prof. in Industrial Chemistry 1936-1938,
whose life was devoted to science, work, respect and justice.

\end{document}